# Novel Self-passivation Rule and Structure of CdTe ∑3 (112) Grain Boundary

Cheng-yan Liu[1,2], Yue-yu Zhang[1,2], Yu-sheng Hou[1,2], Shi-you Chen[3], Hong-jun Xiang[1,2], Xin-gao Gong[1,2]

[1]*Key Laboratory for Computational Physical Science (Ministry of Education), State Key Laboratory of Surface Physics and Department of Physics, Fudan University, Shanghai 200433, China*

[2]*Collaborative Innovation Center of Advanced Microstructures, Nanjing 210093, Jiangsu, China*

[3]*Laboratory of Polar Materials and Devices, East China Normal University, Shanghai 200241, China*

**Abstract**

The theoretical study of grain boundaries (GBs) in polycrystalline semiconductors is currently stalemated by their complicated nature, which is difficult to extract from any direct experimental characterization. Usually, coincidence-site-lattice (CSL) models are constructed simply by aligning two symmetric planes, ignoring various possible reconstructions. Here, we propose a general self-passivation rule to determine the low-energy GB reconstruction, and find new configurations for the CdTe ∑3 (112) GBs. First-principles calculations show that it has lower formation energies than the prototype GBs adopted widely in previous studies. Surprisingly, the reconstructed GBs show self-passivated electronic properties without deep-level states in the band gap. Based on the reconstructed configurations, we revisited the influence of $CdCl_2$ post-treatment on the CdTe GBs, and found that the addition of both Cd and Cl atoms in the GB improves the photovoltaic properties by promoting self-passivation and inducing n-type levels, respectively. The present study provides a new route for further studies of GBs in covalent polycrystalline semiconductors and also highlights that previous studies on the GBs of multinary semiconductors which are based on the unreconstructed prototype GB models, should be revisited.

## I. INTRODUCTION

Polycrystalline semiconductor thin films, which are widely used for solar cells, have the advantages of low-cost and appreciable photovoltaic efficiency. However, the efficiency has always been significantly lower than the theoretical prediction, e.g., the record efficiency of a CdTe solar cells is only 21%,[1] much lower than the theoretical limit of 32% predicted by the Shockley-Queisser model.[2,3] GBs are believed to be responsible for the difference between theoretical and actual efficiency,[4] because they produce deep levels in the band gap and increase the carrier recombination rate significantly.

Determining the GB structures is the first step in the study of GBs, however, it is still challenging both theoretically and experimentally. Simple coincidence-site-lattice (CSL) structure models, which are constructed by aligning two symmetric planes, have been widely used. For example, a CdTe Σ3 (112) GB model (hereafter called a prototype GB) was constructed in this way and adopted in a series of previous theoretical studies.[5-9] It is well known that covalent bonds have prominent reconstruction characteristics, for example, surface reconstructions,[10-13] edge reconstructions[14,15] and interface reconstructions[16,17]. Obviously, the atomic configuration of prototype CSL Σ3 (112) GBs, without carefully accounting for bond rearrangement, is questionable. Similar Σ3 (112) GB models were also used in the study of Σ3 (114) GBs of ternary CuInSe$_2$,[18] Σ3 (114) GBs of quaternary Cu$_2$ZnSnSe$_4$[19] and Σ5 (310) GBs of perovskite CH$_3$NH$_3$PbI$_3$.[20] Therefore, a careful investigation of the possible atomic configurations of Σ3 (114) GBs and their properties is urgently needed and of general importance.

In this paper, we propose a self-passivation rule to determine the GB reconstruction in semiconductors, and found two new self-passivated CdTe Σ3 (112) Te-core GB structures (Cd$_i$, Cd$_i$+V$_{Cd}$). Unlike simply aligning two symmetric crystal planes (the prototype CSL GB structure),[5-9] the self-passivated reconstructed GB structures have lower formation energies, and surprisingly they do not have deep levels in the gap. We further studied the effect of CdCl$_2$ post-treatment, which is an important process for improving the photovoltaic efficiency of CdTe thin films. Cl concentrating at GBs was deservedly considered as the most important dopant to cure the GBs' defect levels.[6,21,22,23] We show that the Cl dopant in these new GB structures[21,23] induces n-type levels, contributing to the separation of electron-hole pairs,[22,24,25] while the Cd dopant, whose role was neglected in previous studies, facilitates the formation of self-passivating GBs that enhance the photovoltaic efficiency. These results change the conventional picture about how the CdCl$_2$ post-treatment improves the photovoltaic efficiency of polycrystalline CdTe solar cells, and also manifest that the previous studies on the most stable Σ3 GBs in a series of semiconductors should be revisited.

## II. METHODS AND CALCULATIONS

### A. Self-passivation rule

The CSL model is basically a geometric model, simply aligning two symmetric planes to satisfy the coincidence condition at the boundary, which inherits the local atomic environment of the bulk. For example, Fig. 1a shows the widely adopted CSL model of CdTe Σ3 (112) GB,[5-9] which is constructed with two (112) planes merged together. At the boundary, six atomic sites from the two grains coincide, and a vacancy region is formed between the two surfaces. Since two Te atoms ($Te_1$ and $Te_1'$) are very close to the core of the vacancy region, this GB is also called Te-core Σ3 (112) GB. Such GB structure from CSL models usually has a small strain energy, however, in semiconductor compounds the local chemical bonding energy might be even more important than strain energy. The numerous dangling bonds would make the CSL structure energetically un-favorable. Actually, previous experimental HRTEM image demonstrated this simple CSL GB structure, but it can not clearly depict the internal atomic configuration, especially for the interstitial atoms which might have low contrast.[5] Therefore, theoretical understanding becomes an important method to reveal the GBs structure.

Here, we propose a self-passivation rule and construct a lower energy GB structure than the standard CSL model: the number of anions with full shells of 8-electron must be maximized through the reconstruction including moving or adding (removing) atoms in the GB core. Although the reconstruction could be large, the gain of chemical bonding energy can overcome the increase of strain energy. For example, the Te-core Σ3 (112) CSL model in Fig. 1a has four dangling bonds on the four Te atoms near the core, and 0.5 extra electron should be added to each Te atom to give it an 8-electron full shell state. In order to decrease the number of dangling bonds, we can add one extra Cd atom (with 2 electrons) to this region, then all four Te anions are in the 8-electron full state. We call this reconstructed GB model with an extra Cd the $Cd_i$ model, as shown in Fig. 1c ($Cd_4$ shows the extra Cd atom). By applying the self-passivation rule to the prototype structure (Fig. 1a), one can simply move the $Cd_2'$ between $Te_2$-$Te_2'$ to the position between $Te_1$-$Te_1'$, and construct a $V_{Cd}+Cd_i$ model as shown in Fig. 1b. This model has essentially three dangling bonds, less than those in the CSL model. The $V_{Cd}+Cd_i$ model should also have a lower formation energy.

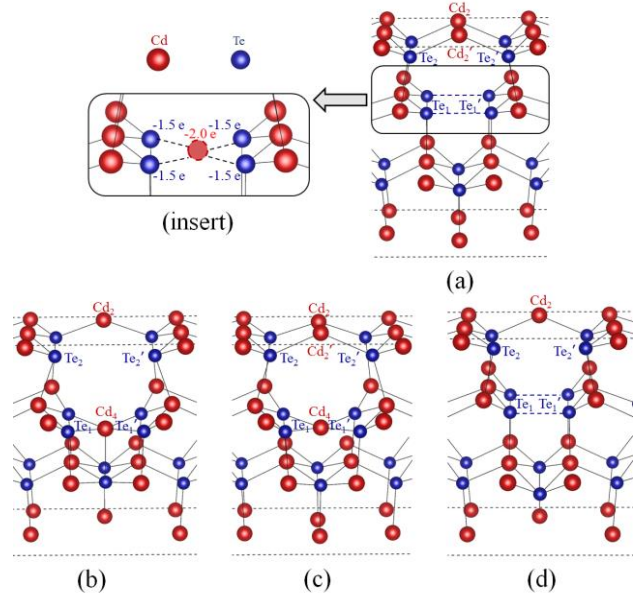

*Fig. 1. Four different kinds of ∑3 (112) GB structures for a Te-core. (a) the prototype Te-core GB constructed by aligning two (112) crystal planes. (b) the $Cd_i+V_{Cd}$ GB constructed by moving the $Cd_2$ atom down into a $Cd_4$ in the center of four $Te_1$ ($Te_1$') atoms. c) the $Cd_i$ GB constructed by adding an interstitial $Cd_4$ atom in the center of four $Te_1$ ($Te_1$') atoms. (d) the $V_{Cd}$ GB formed by removing the $Cd_4$ atom from (b). The (insert) is a local atomic configuration shown in the black frame extracted from the prototype GB (a). Adding a Cd atom in the center of four $Te_1$ ($Te_1$') atoms satisfies the octet electron counting model.*

## B. Details of the calculations

The first-principles calculations were performed based on the density functional theory (DFT) as implemented in the Vienna Ab-initio Simulation Package (VASP).[26,27] The projector-augmented wave (PAW) method with PBE functional[28] was used in the simulations. The aforementioned procedures were also used for the studies of Cd-core GBs in supplement material.[29] In order to eliminate the interaction of internal electric fields between GB models, a slab (31 layers in (112) direction) containing 62 Te atoms and 62 Cd atoms was constructed for single Te-core ∑3 (112) prototype GBs to calculate the individual GBs' properties as shown in Fig. S1 in the supplement material.[29] Surface dangling bonds were passivated with 16 pseudo-hydrogen atoms to mimic the bulk properties.[30] The lattice constant of the super cell is a= 9.27 Å, b=11.43 Å, c=58.68 Å with 15 Å of vacuum separating the slabs. The models are fully relaxed with Hellmann-Feynman forces less than 0.02 eV/Å and electronic convergence less than $1\times10^{-4}$ eV. The energy cut-off is 320 eV and the k points are sampled with a 4×4×1 mesh in the Brillouin zone.[31] To check the quantum confinement effects,[32,33] slab structures from 31 to 55 layers with an interval of 6 layers were examined. The results showed that the band gap for the slab of 31 layers is just ~0.1 eV higher than that in the bulk, which would not severely affect our electronic structure results.

The relative formation energy[34,35] of the doped GBs is defined as

$$\Delta H_f = \Delta E + n_{Cd}\mu_{Cd} + n_{Te}\mu_{Te} + n_{Cl}\mu_{Cl},$$

$$\Delta E = E(defect_{GB}) - E(prototype_{GB}) + n_{Cd}\mu_{Cd}^0 + n_{Te}\mu_{Te}^0 + n_{Cl}\mu_{Cl}^0,$$

where $E(prototype_{GB})$ is the total energy of prototype GB which is set as the reference, $E(defect_{GB})$ is the total energy of the GB doped with Cl or reconstructed GBs. $n_{Cd}$, $n_{Te}$ and $n_{Cl}$ are the number of Cd, Te and Cl atoms referenced to the prototype GB. $\mu_{Cd}$, $\mu_{Te}$ and $\mu_{Cl}$ are the chemical potentials of Cd, Te and Cl relative to the chemical potentials $\mu_{Cd}^0$, $\mu_{Te}^0$ and $\mu_{Cl}^0$ of their elemental phase.

## III. RESULTS AND DISCUSSION

### A. Stable reconstruction GBs

Our first-principles calculations on these models show that the new reconstructed models are energetically more favorable and electronically more benign than the prototype CSL model. Fig. 2a shows the formation energies of the four different kinds of Te-core ∑3 (112) GBs as a function of the Te chemical potential. The region of elemental chemical potentials is thermodynamically limited by several conditions. First, $\mu_{Cd} + \mu_{Te} = \Delta H_{CdTe}$ (−1.08 eV) maintains a stable CdTe phase. Second, $\mu_{Cd} < 0$ and $\mu_{Te} < 0$ avoid the formation of the pure elemental phases of Cd and Te. The formation energy of $Cd_i+V_{Cd}$ GBs is always lower than that of the prototype in the whole chemical potential range, and it has the lowest formation energy in the intermediate region of Te chemical potential (−0.99 eV $< \mu_{Te} <$ −0.3 eV). We have also searched the atomic structure of this GB using the structure evolution searching method which was implemented in the IM²ODE package[36], and arrived at the $Cd_i+V_{Cd}$ structure as the lowest energy structure at the stoichiometric condition.

In the Te poor region (−1.08 eV $< \mu_{Te} <$ −0.99 eV), the $Cd_i$ GB (Fig. 1c) has the lowest formation energy. In order to make a comparison with $Cd_i+V_{Cd}$, we constructed the $V_{Cd}$ structure by removing a $Cd_4$ atom from $Cd_i+V_{Cd}$ as shown in Fig. 1d. We can see that in the Te rich region (−0.30 eV $< \mu_{Te} <$ 0 eV), the $V_{Cd}$ GB structure is stable. For Cd-core GBs which are shown in Fig. S2 of the Supplement Material,[29] the prototype GB, which has no anion dangling bonds, is the most stable GB structure in the entire chemical potential range, based on the formation energy studies shown in Fig. S3a.

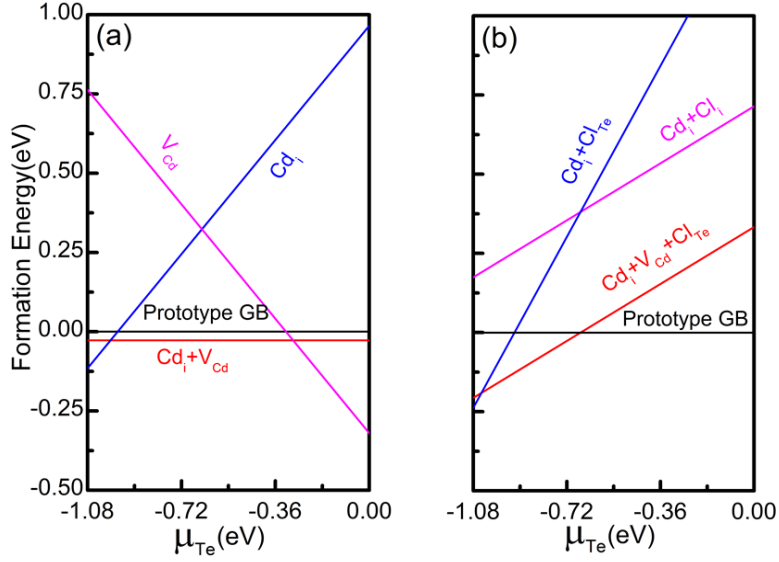

*Fig. 2. (a) The formation energies of the Te-core GB structures. The $V_{Cd}$, $Cd_i+V_{Cd}$ and $Cd_i$ GBs have lower formation energies than the prototype GB within certain ranges of the chemical potential. (b) The formation energies of the Cl-dopant in self-passivated Te-core GBs. $Cd_i+Cl_{Te}$, $Cd_i+Cl_i$ and $Cd_i+V_{Cd}+Cl_{Te}$ GBs are calculated with Cl-doped in newly proposed GBs ($Cd_i$ and $Cd_i+V_{Cd}$).*

## B. Benign electronic properties of reconstruction GBs

From the electronic structure calculations of the self-passivated $Cd_i+V_{Cd}$ and $Cd_i$ GBs (Figs, 3c and 3d), we find no deep-level states in the gap. The deep levels are localized at the center of the band gap for both the prototype and $V_{Cd}$ GB as shown in Figs. 3a and 3b, respectively. To clearly identify the origin of the deep levels, we studied the local density of states (LDOS) in the vicinity of GBs as is shown in Fig. 3. The LDOS shows that two wrong bonds formed by $Te_1$-$Te_1$' contribute to the localized deep levels in the gap, in agreement with previous studies.[7,8] Based on the octet electron rule,[37] both $Te_1$ and $Te_1$' are triply coordinated with $-6+3\times(-2/4)=-7.5$ electrons. Therefore, the $Te_1$-$Te_1$' bond (3.56 Å long) forms, sharing 0.50 electron to satisfy the local octet rule, and creating shallow bonding states hidden in the valence bands (VB) and anti-bonding states acting as the deep levels in the gap.

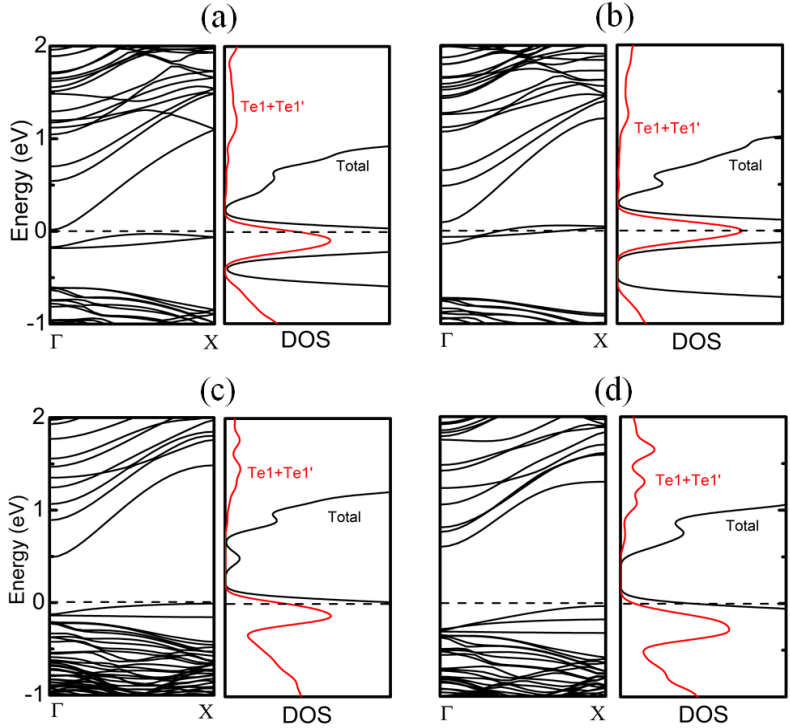

*Fig. 3. Band structures and LDOS for (a) prototype GB, (b) $V_{Cd}$ GB, (c) $Cd_i+V_{Cd}$ GB and (d) $Cd_i$ GB for Te-core GBs. The dashed lines indicate the energy positions of the highest occupied states. The prototype and $V_{Cd}$ GB have similar electronic structures with the localized deep defect levels created by $Te_1$-$Te_1$' bonds. $Cd_i+V_{Cd}$ and $Cd_i$ GBs do not have the deep-defect levels since two pairs of $Te_1$-$Te_1$' wrong bonds are replaced by the four $Cd_4$-$Te_1$ ($Te_1$') bonds.*

The band structure and LDOS of the stoichiometric $Cd_i+V_{Cd}$ GB (Fig. 3c) clearly show that the defect levels have moved below the valence band maximum (VBM), leaving a relatively large band gap. The reason is that, as soon as the $Cd_4$ atom is inserted into the center of four $Te_1$ ($Te_1$') atoms, instead of two pairs of $Te_1$-$Te_1$' wrong bonds in the prototype GB or $V_{Cd}$, four Cd-Te normal bonds are formed. Two pairs of $Te_1$-$Te_1$' wrong bonds are broken with a distance of 4.00 Å. Comparing with prototype GB, the net effect of $V_{Cd}+Cd_i$ is that four dangling bonds of $Te_1$-$Te_1$ atoms are replaced by four $Cd_4$-$Te_1$ ($Te_1$') normal bonds as well as leaving two dangling bonds on $Te_2$ and $Te_2$' atoms. It is well known that the tetra-coordinated Te-Cd bond has lower bonding states in the VB and higher anti-bonding states in the conduction band (CB) than that of Te-Te bonds. Therefore, the deep level states are eliminated. For the $Cd_i$ GB at Te poor condition, the deep levels are further eliminated by reducing more dangling bonds compared to Te poor $Cd_i+V_{Cd}$, as shown in Fig. 3d. Therefore, the self-passivated GBs ($V_{Cd}+Cd_i$ and $Cd_i$) not only have relatively lower formation energies in certain chemical potential ranges, but also exhibit benign electronic properties in CdTe thin films.

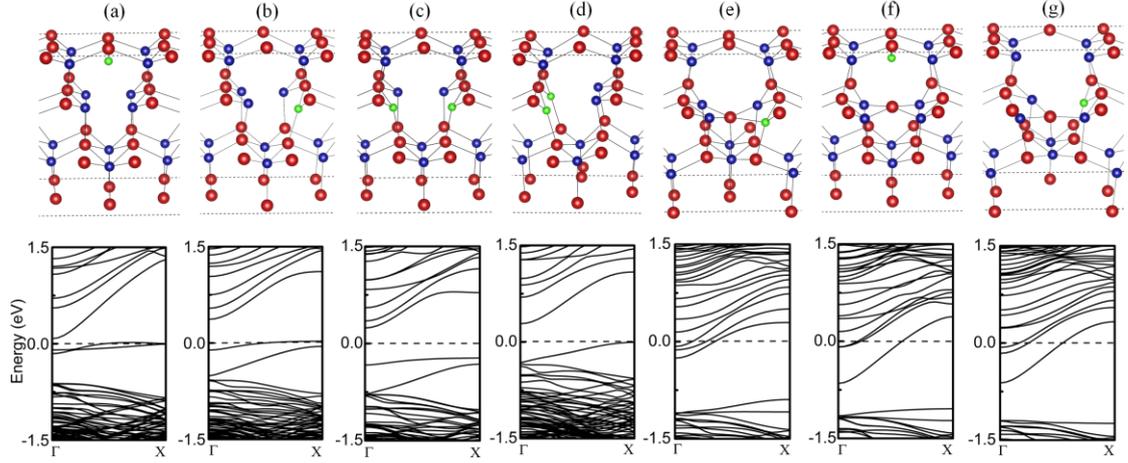

*Fig. 4. The band structures and GB structures of the Cl dopant in Te-core GBs. Prototype GB with Cl dopant for (a) $Cl_i$, (b) $Cl_{Te}$, (c) $2Cl_{Te}$ (in reference 6) and (d) $2Cl_{Te}$'. Reconstructed GB with Cl dopant for (e) $Cd_i+Cl_{Te}$ GB, (f) $Cd_i+Cl_i$ GB, (g) $Cd_i+V_{Cd}+Cl_{Te}$ GB. The dashed lines at zero energy in (a)-(g) indicate the positions of the highest occupied states. Configurations (c) and (d) are semiconductors, whereas (e)-(g) are n-type. The red, blue and green atoms represent Cd, Te and Cl atoms, respectively.*

### C. Novel understanding on CdCl$_2$ post-treatment

CdCl$_2$ post-treatment was considered as the best approach to improve the photovoltaic efficiency in experiments.[23,25,38] Previous studies based on the prototype GBs showed that only the Cl from CdCl$_2$ post-treatment contributed to the improvement, because $Cl_{Te}$ eliminated the deep-level states induced by the GBs in the gap.[6] However, the present study clearly shows that the prototype structure is not energetically the most favorable, so the mechanism of CdCl$_2$ post-treatment should be revisited.

We verified that in the prototype GB, only $2Cl_{Te}$ and $2Cl_{Te}$' (Fig. 4c and 4d) can eliminate the deep levels. However, if two Cl atoms replace two Te$_1$ (or two Te$_2$) atoms as $2Cl_{Te}$' in Fig. 4d, which breaks two Te$_1$-Te$_1$' wrong bonds, the $2Cl_{Te}$' will result in lower formation energy than that of $2Cl_{Te}$ (Fig. 4c) reported in Ref. 6, in which two Cl atoms are replaced by one Te$_1$ and one Te$_1$'. Unavoidably, one Cl with the formation of $Cl_i$ and $Cl_{Te}$ (Figs. 4a and 4b), still leaves the deep-defect levels in the gap of prototype GB.

The Cl dopant was found to induce n-type GBs in the self-passivated $Cd_i$ and $V_{Cd}+Cd_i$ GBs, which can promote the separation of electron-hole pairs. Fig. 2b gives the formation energies for Cl doped self-passivated structures corresponding to Figs. 4e, 4f and 4g. In those figures, with the $Cd_i+Cl_{Te}$, $Cd_i+Cl_i$ and $V_{Cd}+Cd_i+Cl_{Te}$ GBs, one can see that Cl donates electrons to the CB and pushes the highest occupied states down in the VB. The reason is that self-passivated GBs ($Cd_i$ and $V_{Cd}+Cd_i$) already saturate the electron charge of the two pairs of Te$_1$-Te$_1$' wrong bonds without any deep-defect states and Cl doping increases the number of valence electrons in the system by the formation of $Cl_i$ and $Cl_{Te}$.

The effect of Cd addition by CdCl$_2$ post-treatment in Cd rich conditions had been completely neglected in previous studies. Our present results indicate that Cd addition facilitates the prototype GB and V$_{Cd}$ transition into the self-passivated GBs (Cd$_i$ and V$_{Cd}$+Cd$_i$) which exhibit a benign band gap. For Cd-core GBs, detailed investigations with Cl dopants are shown in S3b (formation energy) and S4 (GB structures and energy bands). The most stable structure in Cd-core GB, in the whole chemical potential range with benign electronic properties, is 2Cl$_i$. So, the present study reveals two positive effects of CdCl$_2$ post-treatment on CdTe GBs: (1) Cd rich conditions can make the Te-core GBs trend to self-passivated GBs (Cd$_i$ and V$_{Cd}$+Cd$_i$) and (2) Cl doping in the self-passivated GBs (Cd$_i$ and V$_{Cd}$+Cd$_i$) makes the GBs inverted into n-type, forming p-n-p junctions buried in the grains. These effects are in contrast with the previous understanding based on the prototype GBs.

## IV. SUMMARY

We have proposed a general GB self-passivation rule for constructing the GB structures. We find that not only do the self-passivated GBs (Cd$_i$ and V$_{Cd}$+Cd$_i$) have lower formation energies compared to the prototype GB proposed before, but also a band gap is opened. With the self-passivated GB structures, we re-examined the effect of CdCl$_2$ post-treatment, and revealed a novel mechanism for how the CdCl$_2$ post-treatment improves the photovoltaic performance. The addition of Cl inverts the GBs into n-type in self-passivated GBs with the formation of Cl doping (Cd$_i$+Cl$_{Te}$, Cd$_i$+Cl$_i$ and Cd$_i$+V$_{Cd}$+Cl$_{Te}$) instead of passivating the deep levels of the prototype GB (2Cl$_{Te}$). The effect of Cd addition, which was neglected previously, facilitates the formation of self-passivating GBs. The present study sheds light on understanding the atomic structures and electronic properties of GBs in polycrystalline semiconductors. In fact, we have successfully applied this rule to study the properties of GBs in CuInSe$_2$ and Cu$_2$ZnSnS$_4$ [39].

**Acknowledgements**

We greatly thank the stimulating discussion with S.-H. Wei. This work was partially supported by the Special Funds for Major State Basic Research, National Science Foundation of China (NSFC), international collaboration project of MOST, Pujiang plan, Program for Professor of Special Appointment (Eastern Scholar) and Shanghai Rising-star program. Computation was performed in the Supercomputer Center of Fudan University.

**References**

[1]M. A. Green, K. Emery, Y. Hishikawa, W. Warta, E. D. Dunlop, Solar cell efficiency tables (version 46). Prog. Photovolt: Res. Appl. **23**, 805, (2015).

[2]W. Shockley, H. J. Queisser, J. Appl. Phys., **32**, 510, (1961).

[3]M. A. Green, Prog. Photovolt: Res. Appl., **20**, 472, (2012).

[4]A. P. Sutton, R. W. Balluffi, Interfaces in Crystalline Materials (Oxford Science Publications, New York, 1995).

[5]Y. Yan, M. M. Al-Jassim, K. M. Jones, J. Appl. Phys., **94**, 2976, (2003).

[6]L. Zhang, J. L. F. Da Silva, J. Li, Y. Yan, T. A. Gessert, S.-H. Wei, Phys. Rev. Lett. , **101**, 155501, (2008).

[7]Y. Yan, M. M. Al-Jassim, K. M. Jones, J. Appl. Phys., **96**, 320, (2004).

[8]C. Feng, W.-J. Yin, J. Nie, X. Zu, M. N. Huda, S.-H. Wei, M. M. Al-Jassim, Y. Yan, Solid State Commun., **152**, 1744, (2012).

[9]J.-S. Park, J. Kang, J.-H. Yang, W. Metzger, S.-H. Wei, New J. Phys. **17**, 013027, (2015).

[10]Y. P. Feng, T. H. Wee, C. K. Ong, H. C. Poon, Phys. Rev. B, **54**, 7, (1996).

[11]J. E. Bickel, N. A. Modine, C. Pearson, J. M. Millunchick, Phys. Rev. B, **77**, 125308, (2008).

[12]S. Iarlori, G. Galli, F. Gygi, M. Parrinello, E. Tosatti, Phys. Rev. Lett., **69**, 20, (1992).

[13]Q. Wang, A. R. Oganov, Q. Zhu, X. F. Zhou, Phys. Rev. Lett., **113**, 266101, (2014).

[14]P. Koshinen, S. Malola, H. Häkkinen, Phys. Rev. Lett., **101**, 115502, (2008).

[15]M. C. Lucking, J. Bang, H. Terrones, Y. Y. Sun, S. Zhang, Chem. Mater. **27**, 3326, (2015).

[16]H. Hibino, T. Ogino, Phys. Rev. B, **49**, 8, (1994).

[17]R. L. Headrick, A. F. J. Levi, H. S. Luftman, J. Kovalchick, L. C. Feldman, Phys. Rev. B, **43**, 18, (1991).

[18]Y. Yan, C.-S. Jiang, R. Noufi, S.-H. Wei, H. R. Moutinho, M. M. Al-Jassim, Phys. Rev. Lett., **99**, 235504, (2007).

[19]W.-J. Yin, Y. Wu, S.-H. Wei, R. Noufi, M. M. Al-Jassim, Y. Yan, Adv. Eng. Mater., **4**, 1300712, (2014).

[20]W.-J. Yin, H. Chen, T. Shi, S.-H. Wei, Y. Yan, Adv. Electron. Mater., **1**,1500044, (2015).

[21]C. Li, Y. Wu, J. Poplawsky, T. J. Pennycook, N. Paudel, W. Yin, S. J. Haigh, M. P.

Oxley, A. R. Lupini, M. Al-Jassim, S. J. Pennycook, Y. Yan, Phys. Rev. Lett., **112**, 156103, (2014).

[22]I. V. Fisher, S. R. Cohen, A. Ruzin, D. Cahen, Adv. Mater., **16**, 11, (2004).

[23]D. Mao, C. E. Wickersham, Jr., M. Gloeckler, IEEE J. Photovoltaics, **4**, 6, (2014).

[24]R. W. Birkmire, E. Eser, Annu. Rev. Mater. Sci., **27**, 625, (1997).

[25]R. W. Birkmire, B. E. McCandless, Curr. Opin. Solid State Mater. Sci., **14**, 139, (2010).

[26]G. Kresse, J. Hafner, Phys. Rev. B, **47**, 558, (1993).

[27]G. Kresse, J. Hafner, Phys. Rev. B, **49**, 14251, (1994).

[28]G. Kresse, D. Joubert, Phys. Rev. B, **59**, 1758, (1999).

[29]See Supplemental Material at URL for the properties of prototype Cd-core $\sum 3$ (112) GB doped with Cl.

[30]H. X. Deng, S. S. Li, J. Li, S.-H. Wei, Phys. Rev. B, **85**, 195328, (2012)

[31]H. J. Monkhorst, J. D. Pack, Phys. Rev. B, **13**, 5188, (1976).

[32]J. W. Haus, H. S. Zhou, I. Honma, H. Komiyama, Phys. Rev. B, **47**, 1359, (1993)

[33]M. V. R. Krishana, R. A. Friesner, J. Chem. Phys., **95**, 8309, (1991)

[34]S. Chen, J. H. Yang, X. G. Gong, A. Walsh, S.-H. Wei, Phys. Rev. B, **81**, 245204, (2010)

[35]B. Liu, Z. Qi, C. Shi, Phys. Rev. B, **74**, 174101, (2006)

[36]Y. Zhang, W. Gao, S. Chen, H. Xiang, X. Gong, Comput. Mater. Sci., **98**, 51, (2015).

[37]L. Zhang, E. G. Wang, Q. K. Xue, S. B. Zhang, Z. Zhang, Phys. Rev. Lett., **97**, 126103, (2006).

[38]R. E. Treharne, B. L. Williams, L. Bowen, B. G. Mendis, K. Durose, in Photovoltaic Specialists Conference (PVSC) 38th IEEE, Austin, TX, 2012 (IEEE, Singapore, 2012) doi:10.1109/PVSC.2012.6317985 pp. 001983–001987.

[39]Chen-Yan Liu, Zhi-ming Li, Hong-yang Gu, Shi-you Chen, Hong-jun Xiang, Xin-gao Gong, to be submitted.